\documentclass[]{article}

\usepackage{slashed}
\newcommand{\bb}{\begin{eqnarray}}
\newcommand{\ee}{\end{eqnarray}}
\begin{document}

\title{Axion field induces exact symmetry}
\author{P. Mitra\footnote{parthasarathi.mitra@saha.ac.in}\\
Saha Institute of Nuclear Physics,\\
1/AF Bidhannagar\\ Calcutta 700064} 
\date{}
\maketitle

\begin{abstract}
While no regularization is consistent with the anomalous chiral symmetry which
occurs for massless fermions, 
the artificial axion-induced symmetry for massive fermions 
is shown here to be consistent with a standard regularization, 
even in curved spacetime,
so that it can be said to have no anomaly in gauge or gravitational fields.
Implications for $\theta$ terms are pointed out.
\end{abstract}

\section{Introduction}
Chiral symmetry, which is exact for massless fermions at the classical level
and approximate for light quarks, has been very useful in particle physics.
In analogy with this chiral symmetry,
an artificial chiral-like symmetry [1] was introduced 
some time back for the strong interactions with massive quarks.
It implies the occurrence of a light
pseudoscalar particle, the axion [2], which has however not been detected in
spite of elaborate searches [3]. 
This symmetry has been suspected to be fraught with an anomaly because of the 
involvement of a chiral transformation. However, this is not obvious
and has to be examined by careful regularization of the theory.
This is important because of implications for the symmetry of the theory.
 
While chiral symmetry of the action, corresponding to the transformation
\bb
\psi\rightarrow e^{i\alpha\gamma_5}\psi,\quad
\bar\psi\rightarrow \bar\psi e^{i\alpha\gamma_5},
\ee
is broken by a non-vanishing quark mass term $m\bar\psi\psi$, with mass $m$,
the artificial chiral symmetry for massive fermions works by letting a new field
${\varphi}$ absorb the chiral transformation. The mass term is replaced
by
\bb\bar\psi m e^{i{\varphi}\gamma_5}\psi,\ee
which is classically invariant under the transformation
\bb
\psi\rightarrow e^{i\alpha\gamma_5}\psi,\quad
\bar\psi\rightarrow \bar\psi e^{i\alpha\gamma_5},
\quad\varphi\rightarrow{\varphi}-2\alpha,
\label{0}\ee
which is also a symmetry of the other terms of the action. 

The original interaction introduced by Peccei and Quinn [1] was of the form
\bb
\bar\psi [\Phi{1+\gamma^5\over 2}
+ \Phi^\dagger{1-\gamma^5\over 2}]\psi,\ee
where $\Phi$ is a complex scalar field with a symmetry breaking potential.
The artificial chiral symmetry transformation here is
\bb\Phi\to e^{-2i\alpha}\Phi,\quad
\psi\to e^{i\alpha\gamma_5}\psi,
\quad\bar\psi\to \bar\psi e^{i\alpha\gamma_5}.\ee
$\Phi$ may be taken to be of the form $\rho e^{i{\varphi}}$.
The amplitude $\rho$ of the scalar field acquires a vacuum expectation value
because of symmetry breaking, which provides a massive boson.
The phase ${\varphi}$ is the zero mode of the potential
and provides a Goldstone boson.
This is the axion, which should acquire
a mass because of the quark masses, but does not appear to exist.
The kinetic term of $\Phi$ yields 
$\frac12\rho_0^2\partial_\mu\varphi\partial^\mu\varphi$ 
as the kinetic term for $\varphi$, with $\rho_0$ the vacuum expectation
value of $\rho$.
The mass $m$ comes from $\rho_0$ and the coupling constant of $\Phi$.

The anomaly, which occurs for the usual chiral symmetry
and makes the divergence of the axial current $\bar\psi\gamma_\mu\gamma_5\psi$
nonvanishing,
appears when the singularities in the theory are handled by regularization.
Even the measure picture
of anomalies requires a regularization for actual calculations.
We shall therefore use an explicit
regularization for studying chiral transformations of fermions
in the presence of the axion coupling. This will tell us whether
the continuous symmetry (\ref{0})
introduced classically for massive quarks is anomalous
or not. 

%\section*{Symmetries}

The new symmetry is expected to be broken
spontaneously by the vacuum because of the translation of the spinless field
by a c-number. The field must have a definite value in a fixed vacuum:
let it be denoted by $\varphi_0$. Then
\bb
\varphi'=\varphi-\varphi_0
\ee
is the shifted field with vanishing vacuum expectation value. This is a
pseudoscalar field, so that $\bar\psi\varphi'\gamma_5\psi$ and
$(\varphi' {\rm ~tr~}F^{\mu\nu}\tilde F_{\mu\nu})$ are scalars. On the other
hand, $(\varphi_0 {\rm ~tr~}F^{\mu\nu}\tilde F_{\mu\nu})$ is
a pseudoscalar,
like the $\theta$ term $(\theta {\rm ~tr~}F^{\mu\nu}\tilde F_{\mu\nu})$.
If a $(\varphi_0 {\rm ~tr~}F^{\mu\nu}\tilde F_{\mu\nu})$ term is generated
when the fermion is integrated out, it
can modify the $\theta$ term. The possibility of this occurrence is related to
the question of an anomaly in the symmetry (\ref{0}).

If the symmetry (\ref{0}) is anomalous,
no regularization of the fermions will be consistent with
it. On the other hand, if any regularization is found to be consistent with
it, the symmetry has no anomaly in this quantization and
such symmetric regularizations are the ones to be preferred.
Pauli-Villars regularization will be studied in flat and in curved spacetime.

\section{Pauli-Villars regularization}
We recall first the simple form of the Pauli-Villars
regularization. For a fermion with mass $m$, the Lagrangian density 
\bb
\bar\psi[i \slashed{D} -m]\psi
+\bar\chi[i \slashed{D} -M]\chi
\ee
involves regulator spinor fields $\chi,\bar\chi$ which however are
assigned Bose statistics. The regulator mass $M$ is ultimately taken to
infinity when the regulator fields decouple. The regulator fields couple to the
gauge fields in the standard way. The regulator mass term breaks the
chiral symmetry and yields the chiral anomaly in the $M\to\infty$ limit.
This is the original form of the Pauli-Villars regularization. A more general
form with several species of regulator fields is also available [4]:
\bb
\bar\psi[i \slashed{D} -m]\psi
+\sum_j\sum_{k=1}^{|c_j|}\bar\chi_{jk}[i \slashed{D} -M_j]\chi_{jk}.
\ee
Here $c_j$ are integers whose signs are related to the statistics assigned
to $\chi_{jk}$. They have to satisfy some conditions to ensure regularization
of the divergences [4]:
\bb
1+\sum_jc_j=0,\quad m^2+\sum_jc_jM_j^2=0.
\ee

Chiral transformations work on $\chi$ as well as $\psi$. Hence
the axion coupling has to be introduced for both $\psi,\chi$,
like the gauge coupling. One has
\bb
\bar\psi[i \slashed{D} -me^{i\varphi\gamma_5}]\psi
+\bar\chi[i \slashed{D} -Me^{i\varphi\gamma_5}]\chi.
\label{4}\ee
This regularization (\ref{4}) is invariant under the combined transformation
\bb
\psi\rightarrow e^{i\alpha\gamma_5}\psi,&\quad&
\bar\psi\rightarrow \bar\psi e^{i\alpha\gamma_5},\nonumber\\
\chi\rightarrow e^{i\alpha\gamma_5}\chi,&\quad&
\bar\chi\rightarrow \bar\chi e^{i\alpha\gamma_5},\nonumber\\
\varphi&\rightarrow&{\varphi}-2\alpha,
\ee
which is the Pauli-Villars extension of (\ref{0}).
This means the symmetry survives when the regularization is
removed by taking the limit $M\to\infty$
and hence is not anomalous. Unlike the regularized axial current, which has
a pseudoscalar divergence arising from the masses,
the Noether current for the extended symmetry, namely
\bb
\bar\psi\gamma_\mu\gamma_5\psi+\bar\chi\gamma_\mu\gamma_5\chi
+2\rho_0^2\partial\mu\varphi,
\ee
is conserved.
The non-vanishing divergence of the axial current is cancelled by that
of the $\varphi$ piece by virtue of the axion equation of motion.

It is interesting to note that
the $\varphi$ phases in (\ref{4}) can be removed by a joint chiral
transformation of both the physical fermions and the regulator spinors:
\bb
\psi\to e^{-i\varphi\gamma_5/2}\psi,&&
\bar\psi\to \bar\psi e^{-i\varphi\gamma_5/2},\nonumber\\
\chi\to e^{-i\varphi\gamma_5/2}\chi,&&
\bar\chi\to \bar\chi e^{-i\varphi\gamma_5/2}.
\ee
The derivative operators in the action produce derivatives of $\varphi$,  
which appear after the transformation.
Now the transformation has a trivial Jacobian because the contribution of the
fermion field is cancelled by that of the bosonic
regulator field as in the latter case the determinant arising from
functional integration comes in the denominator. The argument for the
generalized Pauli-Villars regularization, with (\ref{4}) replaced by
\bb
\bar\psi[i \slashed{D} -me^{i\varphi\gamma_5}]\psi
+\sum_j\sum_{k=1}^{|c_j|}\bar\chi_{jk}[i \slashed{D} -M_je^{i\varphi\gamma_5}]\chi_{jk},
\ee
involves a Jacobian with the factor $1+\sum_jc_j=0$
in the exponent [5]. Consequently the effective action does not contain
any $\varphi {\rm ~tr~}F^{\mu\nu}\tilde F_{\mu\nu}$ in this regularization.
The conserved Noether current here is
\bb
\bar\psi\gamma_\mu\gamma_5\psi+\sum_j\sum_{k=1}^{|c_j|}
\bar\chi_{jk}\gamma_\mu\gamma_5\chi_{jk}+2\rho_0^2\partial\mu\varphi,
\ee

Note that the argument may be easily extended to curved spacetime,
where the Dirac operator
\bb
i\slashed{D}=i\gamma^le^\mu_l(\partial_\mu-iA_\mu-\frac{i}{2}A^{mn}_\mu\sigma_{mn})
\ee
comes with a tetrad $e^\mu_l$ and a spin connection $A^{mn}_\mu$ in
addition to the gauge field, but
it continues to anticommute with $\gamma^5$.
The regularized action is again invariant under the Pauli-Villars extension of 
(\ref{0}).
This means that there is not even an $R\tilde R$ anomaly [6] in this symmetry.
Similar results have also been found by others [7].
%%%%%%%%%%%%%%%%%%%%%%%%%%%%%%%%%%%%%%%%%%%%%%%%%%%%%%%%%%%%%%%%%%%%%%%
\section{Conclusion}
\bigskip

We have examined the chiral symmetry of the fermion action including
$\varphi$. This classical symmetry is preserved by the explicit regularization
(\ref{4}). 
When an acceptable regularization preserves a symmetry, one can 
conclude that the quantum theory defined by the limit of that regularization
will also satisfy it. Hence, after the
fermion is integrated out, the effective action must be invariant under
translations of $\varphi_0$, {\it i.e.,} independent of this variable.
This is indeed true of this regularization, as seen above, and there
is no $\varphi {\rm ~tr~}F^{\mu\nu}\tilde F_{\mu\nu}$ term. This makes the 
modification of a $\theta$ term by a contribution from $\varphi_0$ impossible.
The field $\varphi$ occurs only in the form of derivatives.
The extension of the argument to curved spacetime means that an existing
$R\tilde R$ term [8] too cannot be modified.

There is another way of seeing this.
The exact classical chiral symmetry could be expected to play a r\^{o}le like
the chiral symmetry that holds for massless fermions. With massless fermions,
the $\theta$ term can be modified by a chiral transformation which does not
alter the fermion action but produces an $F\tilde F$ term because of the chiral 
anomaly. However,
the chiral symmetry which occurs in the presence of mass and $\varphi$ is not
anomalous, as we have seen.
So $\theta$ cannot be modified in that way.

It should be pointed out here that any
effect of $\theta$ terms can be removed by setting $\theta=0$ [5] or by making 
it explicitly dynamical, which forces the topological charge to vanish.
A dynamical $\theta$-like term is essentially a Peccei-Quinn 
$(\varphi_0 {\rm ~tr~}F^{\mu\nu}\tilde F_{\mu\nu})$ term without
an axion particle. However, as only fields are dynamical in field theory,
the field $\varphi$ was used in [1] and a direct
$(\varphi {\rm ~tr~}F^{\mu\nu}\tilde F_{\mu\nu})$ term can remove $\theta$.

One may wonder how much freedom one has in choosing regularizations.
Different regularizations have been used in the past
and it is known that all do not lead to the same result. A key point is that
symmetries of the action are generally sought to be preserved by the
regularization. Thus one is always looking for Lorentz invariant and gauge
invariant regularizations for actions having such symmetries.
Regularizations which maintain all symmetries are technically natural.
We have seen that 
the Pauli-Villars regularization
conveniently respects the symmetry (\ref{0}) of the fermion action. 
Non-symmetric regularizations are possible, but are not 
preferable in any way. 
Regularizations that break an
existing classical symmetry like (\ref{0}) may to that extent be
termed technically unnatural. 
The artificial chiral symmetry here is not anomalous
as a regularization consistent with it has been demonstrated to exist. In these
circumstances, working with regularizations inconsistent with it would be
a needless and avoidable violation.

We end with the hope that the exact chiral symmetry in the presence of axions
will be useful also in other calculations and even in curved spacetime.
%%%%%%%%%%%%%%%%%%%%%%%%%%%%%%%%%%%%%%%%%%%%%%%%%%%%%%%%%%%%%%%%%
%%%%%%%%%%%%%%%%%%%%%%%%%%%%%%%%%%%%%%%%%%%%%%%%%%%%%%%%%%%%%%%%%%%%
\section*{REFERENCES}
\begin{enumerate}
%\vspace{-1.5cm}
\item R. Peccei, \& H. Quinn, 
{\it Phys. Rev. Letters}
{\bf 38}, 1440 (1977); \\
{\it Phys. Rev.} {\bf D16}, 1791 (1977)
\item S. Weinberg,  
{\it Phys. Rev. Letters} {\bf 40}, 223 (1978);
F. Wilczek, {\it Phys. Rev. Letters} {\bf 40}, 279 (1978)
\item C. Patrignani et al., Review of Particle Physics,
{\it Chin. Phys. } {\bf C40}, 100001 (2016)
\item L. D. Faddeev \& A. A. Slavnov, {\it Gauge Fields}, (Benjamin-Cummings, Reading, 1982)
\item H. Banerjee, D. Chatterjee \& P. Mitra,
{\it Phys. Letters} {\bf B573}, 109 (2003);
see also P. Mitra,  {\it Symmetries and symmetry breaking in field theory}, 
CRC Press, Florida (2014)
\item T. Kimura, Prog. Theo. Phys. {\bf 42}, 1191 (1969) 
\item G. Dvali and L. Funcke, arXiv:1608.08969 (2016)
\item S. Deser, M. Duff and C. Isham, Phys. Lett. {\bf B93}, 419 (1980)
\end{enumerate}

\end{document}